\documentclass[useAMS,usenatbib]{mn2e} 
\usepackage{aas_macros}
\usepackage{graphics}
\usepackage[pdftex]{graphicx}
\usepackage{epstopdf}
\usepackage{epsfig}  
\usepackage{natbib} 
\usepackage{dingbat}
\usepackage{float}
\usepackage{amsmath}
\usepackage{times}
\usepackage[varg]{txfonts}
\usepackage{verbatim} 
\usepackage{multirow,bigdelim} 
\usepackage{color}
\usepackage{vmargin}
\setmarginsrb{1.2cm}{1cm}{1.2cm}{1cm}{1cm}{1cm}{1cm}{1cm}

\newcommand{\hmsun}{{\,h^{-1}\rm M}_\odot}

  \makeatletter
    \renewcommand{\paragraph}{\@startsection{paragraph}{4}{\z@}%
      {-3.25ex\@plus -1ex \@minus -.2ex}%
      {1.5ex \@plus .2ex}%
      {\normalfont\small\centering}}
     
    \renewcommand{\subparagraph}{\@startsection{subparagraph}{5}{\z@}%
      {-3.25ex\@plus -1ex \@minus -.2ex}%
      {1.5ex \@plus .2ex}%
      {\normalfont\small\centering}}
    \makeatother

\newcommand{\ginnungagap}{{\sc Ginnungagap}}
\newcommand{\music}{{\sc MUSIC}}

\newcommand{\kms}{{ km~s$^{-1}$}}
\newcommand{\hMpc}{{ \textit{h}$^{-1}$~Mpc}}

\title[Simulated clusters]{Galaxy clusters in local Universe simulations without density constraints: a long uphill struggle}
\author[Sorce]
{{Jenny G. Sorce$^{1,2}$\thanks{E-mail: \text{jenny.sorce@univ-lyon1.fr / jsorce@aip.de}}, 
}\\
$^1$Univ Lyon, Univ Lyon1, Ens de Lyon, CNRS, Centre de Recherche Astrophysique de Lyon UMR5574, F-69230, Saint-Genis-Laval, France\\
$^2$Leibniz-Institut f\"{u}r Astrophysik, An der Sternwarte 16, 14482 Potsdam, Germany\\
}

\begin{document}

\date{}

\pagerange{\pageref{firstpage}--\pageref{lastpage}} \pubyear{2018}

\maketitle

\label{firstpage}

\begin{abstract}
\indent Galaxy clusters are excellent cosmological probes provided that their formation and evolution within the large scale environment are precisely understood. Therefore studies with
simulated galaxy clusters have flourished. However detailed comparisons between simulated and observed clusters and their population - the galaxies - are complicated by the diversity of clusters and their surrounding environment. An original way initiated by Bertschinger as early as 1987, to legitimize the one-to-one comparison exercise down to the details, is to produce simulations constrained to resemble the cluster under study within its large scale environment. Subsequently several methods have emerged to produce simulations that look like the local Universe. This paper highlights one of these methods and its essential steps to get simulations that not only resemble the local Large Scale Structure but also that host the local clusters. It includes a new modeling of the radial peculiar velocity uncertainties to remove the observed correlation between the decreases of the simulated cluster masses and of the amount of data used as constraints with the distance from us. This method has the particularity to use solely radial peculiar velocities as constraints: no additional density constraints are required to get local cluster simulacra. The new resulting simulations host dark matter halos that match the most prominent local clusters such as Coma. Zoom-in simulations of the latter and of a volume larger than the 30\hMpc\ radius inner sphere become now possible to study local clusters and their effects. Mapping the local Sunyaev-Zel'dovich and Sachs-Wolfe effects can follow.

\end{abstract}


\section{Introduction}

Galaxy clusters are unconditional cosmological probes. However, for that purpose their formation and evolution within the large scale environment must be understood precisely. This task is greatly complicated by the diversity of clusters \citep{1988S&T....75...16S} and their surrounding environment that undoubtedly affect them and their constituents: the galaxies. One-to-one comparisons between simulated and observed clusters and their population as well as maps of clusters' effects like the Sunyaev-Zel'dovich \citep{1969Ap&SS...4..301Z} and the Sachs-Wolfe \citep{1967ApJ...147...73S} effects but also gravitational lensing effects obtained numerically must then be considered carefully. 

An original way initiated by Bertschinger as early as 1987 \citep{1987ApJ...323L.103B}, to legitimize the one-to-one comparison exercise down to the details, is to produce simulations constrained to resemble the object under study. Thus, while typical simulations \citep{1981MNRAS.194..503E} cope with cosmic variance by simulating large volumes \citep[e.g.][Euclid's flagship, for a non-exhaustive list]{2012MNRAS.426.2046A,DeusSimulation2012,2014MNRAS.444.1518V,2015MNRAS.446..521S,2016MNRAS.463.3948D,2016MNRAS.463.1797D} that abide by a prior cosmological model, constrained simulations consist generally in smaller volumes, abiding also by a cosmological model, but whose initial conditions have been constrained by observational data to decrease the cosmic variance. Many techniques resulting in multiple studies have been developed since then \citep[e.g.][]{2001ApJS..137....1B,2010MNRAS.406.1007L,2013MNRAS.429L..84K,2016MNRAS.455.2078S,2016MNRAS.457..172L,2016ApJ...831..164W}. These methods involve several steps from the treatment via modeling of the observational constraints be they redshift surveys \citep[e.g.][]{2006AJ....131.1163S,aih11,2011MNRAS.416.2840L,2012ApJS..199...26H} or peculiar velocities \citep[e.g.][]{1992ApJS...81..413M,1997ApJS..109..333W,2001MNRAS.326..375Z,2007ApJS..172..599S,2008ApJ...676..184T,2013AJ....146...86T,2016AJ....152...50T} to the reconstruction of the initial conditions that will result in the object of study. While peculiar velocities constitute more of a challenge for observers than redshifts, the advantage of peculiar velocities over redshift surveys (used as density maps) is that they are correlated on large scales and are highly linear.  Constrained initial conditions can be built either forwardly \citep[e.g][]{2008MNRAS.389..497K,2013MNRAS.432..894J,2013ApJ...772...63W}  or backwardly \citep[e.g.][]{1989ApJ...336L...5B,1990ApJ...364..349D,1999ApJ...520..413Z,1993ApJ...415L...5G,2008MNRAS.383.1292L}. In the former case, the initial density field is sampled from a probability distribution function consisting of a prior and a likelihood given the observational data. In the latter case, further explained in the next section of this paper, the initial density field is directly obtained from a realization of the density field today.

To summarize, the steps to finally get the product ``a constrained simulation'' are numerous and can be computer time demanding. It can thus be quite challenging to orientate oneself in the diversity of methods  (with all the different steps, assumptions made and computer time they might involve) as well as to understand exactly down to which scale and which area (depending on the survey coverage) these simulations are efficient simulacra of what they are forced to resemble. This knowledge is however essential to lead studies and to draw appropriate conclusions.

This paper investigates one of the backwards methods with all the steps it involves to get simulations that resemble the local Universe, down to the cluster scale, without adding any additional density constraints: the observational information comes from the velocity field only (galaxy radial peculiar velocities), no supplementary assumption is made on the mass of the different local clusters. This paper thus introduces additionally a better modeling of the uncertainties attributed to the peculiar velocities to obtain the final product: a simulation of the local Universe that resemble not only its Large Scale Structure but that also hosts dark matter halos sufficiently massive to be considered as local observed cluster counterparts up to at least the distance of the Coma cluster (70~\hMpc). Previous studies indeed revealed that several refinements in the modeling of the data and the techniques are required to finally get simulacra of the clusters when using only (i.e. without assumptions on the mass of the clusters) peculiar velocities as constraints. Each one of the refinements added to the method, as an additional step, has so far permitted getting proper (mass, position) and stable (in each simulation) Virgo \citep{2016MNRAS.455.2078S} and Centaurus \citep{2018MNRAS.tmp..523S} galaxy cluster simulacra. The better modeling of the uncertainties introduced here permits obtaining simulacra for the Coma cluster that are massive enough and a better stability for the Hydra cluster, as well as a better repartition of the local mass in general in the reproduced local Large Scale Structure between the different clusters.

This paper thus starts with a brief summary of the different steps of the method used so far to model the data and the involved techniques. Then, it presents the refinement in modeling the uncertainties attributed to the peculiar velocities when inserted in the technique used to reconstruct the local density and velocity fields. Finally, comparisons with previous studies made without this new modeling are proposed before concluding.

\section{Production of local Universe simulacra}

\subsection{Method for the 1$^{st}$ \& 2$^{nd}$ generations of constrained initial conditions}
This subsection is a reminder of the method so far with a brief description of each step and its utility and impact on the resulting local Universe-like simulations:
\begin{enumerate}
\item Catalog of constraints: The constraints are radial peculiar velocities of galaxies. They are obtained from a catalog of galaxy distance moduli \citep[][]{2013AJ....146...86T}. Using H$_0$=75.2~(=100h)~\kms~Mpc$^{-1}$ \citep[value given in][]{2013AJ....146...86T}, this catalog allows to probe distances as large as 250~\hMpc\ but 50\% (90\%) of the data are within 70~\hMpc\ (160~\hMpc).  It consists of a gathering of more than 8,000 galaxy direct distance estimates from several distance estimators: about 88\% are from the Tully-Fisher \citep{1977A&A....54..661T} and Fundamental Plane \citep{2001MNRAS.321..277C}  relations. The rest is from Cepheids \citep{2001ApJ...553...47F}, Tip of the Red Giant Branch \citep{1993ApJ...417..553L}, Surface Brightness Fluctuation \citep{2001ApJ...546..681T}, supernovae of type Ia \citep{2007ApJ...659..122J} and other miscellaneous methods.

\item Grouping of constraints and derivation of peculiar velocities: The quality of the reconstruction obtained with a \emph{linear} minimum variance estimator depends strongly on the grouping of individual data points into single points to suppress virial motions in high density regions \citep{2017MNRAS.468.1812S}. The grouping must remove all the virial motions without affecting neither galaxies in the field \citep{2017MNRAS.468.1812S} nor regions of infall \citep{2017MNRAS.469.2859S}. Only then the proper masses are recovered for the most nearby clusters such as Centaurus \citep{2018MNRAS.tmp..523S} where the proliferation of distance measurements in these regions might lead to an ``overgrouping''  making galaxies falling onto the clusters too sparse. 

The cluster masses could be artificially increased by adding point-like density constraints but 1) mass estimates of observed clusters are not easily determined, 2) masses would not be recovered with the sole velocity field, consequently statistical studies would be biased by the chosen masses, 3) only selected clusters would have their masses increased. Combining both redshift surveys as density and velocities is challenging: since these two types of surveys are not subject to the same biases, having both surveys predict exactly the same field on a megaparsec scale is difficult.  It is then more interesting to focus on using only one type of surveys or eventually both but at different distances: velocities nearby and redshift surveys further away where peculiar velocity measurements become sparse.

This paper uses only peculiar velocities directly derived from the grouped distances (of groups). The latter are weighted by their uncertainties linked to the distance estimators used to derive them (see equations 1 to 3 in \citet{2017MNRAS.469.2859S} and \citet{2013AJ....146...86T} for detailed explanations on the attribution of uncertainties). Uncertainties on grouped distances are subsequently propagated to derive peculiar velocity uncertainty. This additional parameter, called $\sigma_{NL}$, is determined so that $\chi^2 = c_i \langle c_ic_j\rangle^{-1} c_j$ (quantity to gauge the linearity of the data with c$_i$ and c$_j$ the constraints plus their uncertainties) divided by the number of degrees of freedom (dof, number of constraints) is about 1. When $\chi^2$/dof$\sim$1, the data are then sampling a typical realization of the assumed prior power spectrum model in the linear theory.

\item Minimization of biases: The grouping described above is inefficient if used without a bias minimization scheme \citep{2015MNRAS.450.2644S} that prevents the spurious infall onto the local volume, removes the dependence of the reconstruction on the choice of the local H$_0$ \citep{2017MNRAS.469.2859S} and that permits reconstructing sharper structures. This technique also enabled to produce the first simulations that host stable (position, mass, merging history) and massive (more than 2$\times$10$^{14}$~$\hmsun$) Virgo dark matter halos without density constraints (see discussion in (ii) about these density constraints) and with the ``aggressive'' grouping released with the catalog of distances \citep{2016MNRAS.455.2078S,2016MNRAS.460.2015S}. It also permitted developing the local Group factory \citep{2016MNRAS.458..900C} at the origin of multiple projects. This minimization scheme relies on the theoretical expectation that the distribution of radial peculiar velocities must be a Gaussian \citep{2001MNRAS.322..901S}. It iteratively reduces spurious non-Gaussianities in the radial peculiar velocity distribution, to retroactively derive overall better distance estimates resulting in a minimization of the effects of biases. Uncertainties $\sigma$ attributed to the biased minimized distances (velocities) are of 5\% \citep{2015MNRAS.450.2644S}. 

\item Relocation and replacement of constraints: Constraints observed today must be relocated to the position of their precursors to get proper initial conditions. At redshift zero in the simulation, objects can then be expected to be at the proper observed positions. The cosmic displacement field is derived with the Wiener-Filter technique \citep[a linear minimum variance estimator of the data assuming a prior cosmological model,][]{1995ApJ...449..446Z,1999ApJ...520..413Z}. The constraints are then displaced using the Reverse Zel'dovich Approximation \citep{2013MNRAS.430..888D}.

Additionally, after their relocation, the \emph{noisy and uncertain radial} peculiar velocities must be replaced by their 3D Wiener-Filter reconstructions \citep{2014MNRAS.437.3586S} to ensure that not only do simulations reproduce the local Large Scale Structure but that there exists stable knots (clear overdensities at clusters' locations) in each simulation.

\item Building of initial conditions: peculiar velocity constraints are finally inserted in the constrained realization technique \citep{1991ApJ...380L...5H,1992ApJ...384..448H,1996MNRAS.281...84V} to derive an estimate of the residual between the Wiener-Filter reconstructed field and the model. It restores statistically the missing structures in the reconstructed field (the Wiener-Filter goes to the mean field in absence of data or in presence of very noisy data). It relies on the combination of the constraints and a random realization field. 

\end{enumerate}

The resolution of the initial conditions (obtained after rescaling the fields obtained with the constrained realization technique) is increased by adding small scale features with codes like \ginnungagap \footnote{https://github.com/ginnungagapgroup/ginnungagap}\, and \music\  \citep{2011MNRAS.415.2101H}. These codes also permit producing zoom initial conditions \citep{2001ApJS..137....1B} to study with a high resolution local objects. It consists in keeping the large scale environment at low resolution for its effect on the region of interest while the resolution is increased solely in this region.\\

\begin{table*}
\begin{center} 
\begin{tabular}{l@{ }@{ }l@{ }@{ }l@{ }@{ }l@{ }@{ }l@{ }@{ }l@{ }@{ }l}
\hline \hline
Generation &\multicolumn{6}{c}{Steps in the method}\\
 & (i)     &   (ii)     &   (iii)  &     (iv)~a   & (iv)~b  & (v)\\
 & Catalog  & Grouping & Bias Minimization & Reverse Zel'dovich   & 3D & Constrained realizations  \\
 &  (Tully et al. 2013)  &  &  & (Doumler et al. 2013) &  (Sorce et al. 2014) & (Hoffman \& Ribak 1991)\\
\hline
1$^{st}$ (Sorce et al. 2016a)  & \checkmark & Tully et al. 2013 & Sorce 2015   &  \checkmark  &  \checkmark  & \checkmark   \\
2$^{nd}$ (Sorce \& Tempel 2018) & \checkmark & Sorce \& Tempel 2017 &  Sorce 2015 &  \checkmark &  \checkmark  &  \checkmark  \\
3$^{rd}$ (This paper)  & \checkmark  & Sorce \& Tempel 2017 & Sorce 2015 &  \checkmark &  \checkmark &  \checkmark  \\
        &					& 		&    + uncertainty modeling & & &\\
        &					&		&   (this paper) & \\
\hline
importance &  & \multicolumn{2}{c}{nearby masses, stability \hspace{1cm} \hspace{1cm}} & positions & stability, knots &  \\
		&	&		& + farther masses & & &\\
\hline
\hline
\end{tabular}
\end{center}
\vspace{-0.25cm}
\caption{Overview of the 3 generations of constrained simulations. The steps of the method are reminded and the differences between the 3 generations are highlighted. The last line gives the importance or impact of the step on the resulting dark matter halo simulacra. The local Large Scale Structure is simulated in all the cases.}
\label{Tbl:1}
\end{table*}

The 1$^{st}$ and 2$^{nd}$ generations of simulations constrained with the observational data from \citet{2013AJ....146...86T} were obtained with these steps.  While the 1$^{st}$ generation was based on the groups from the literature released with the catalog, the 2$^{nd}$ generation used groups obtained with a Friends of Friends (FoF) percolation method, where different linking lengths in radial (along the line of sight) and in transversal (in the plane of the sky) directions are used. Additionally,  the groups are refined using multimodality analysis to separate nearby/merging systems \citep{2016A&A...588A..14T}. The simulations resemble the local Large Scale Structure \citep{2016MNRAS.455.2078S} and host Virgo \citep[1$^{st}$,][]{2016MNRAS.460.2015S} and Centaurus simulacra \citep[2$^{nd}$][]{2018MNRAS.tmp..523S}. 

The bias minimization was essential in getting such results. Without it, the 2$^{nd}$ generation would produce simulations where the success rate in hosting a Virgo simulacrum is only 60\% against 100\% and halo masses would barely reach 10$^{14}~\hmsun$. As for Centaurus, the majority of the halo masses would barely be above 10$^{14}~\hmsun$ and only a few would reach 4$\times$10$^{14}~\hmsun$, still below recent estimates.

In addition, although a dark matter halo simulacrum is present in most if not all the simulations for local clusters further away than these two very nearby clusters. Their masses are barely above 1.5$\times$10$^{14}~\hmsun$. This is certainly in agreement with observational estimates for clusters like Hydra but not for the Coma cluster. \\

As discussed previously, there are severe disadvantages in increasing locally the masses of these clusters. Consistency in using the sole peculiar velocities as constraints is thus highly preferable to preserve the interest in using these simulations for statistical purposes. 

\subsection{Modeling the uncertainties: 3$^{rd}$ generation of constrained initial conditions}

\begin{figure}
\vspace{-2cm}

\includegraphics[width=0.51 \textwidth]{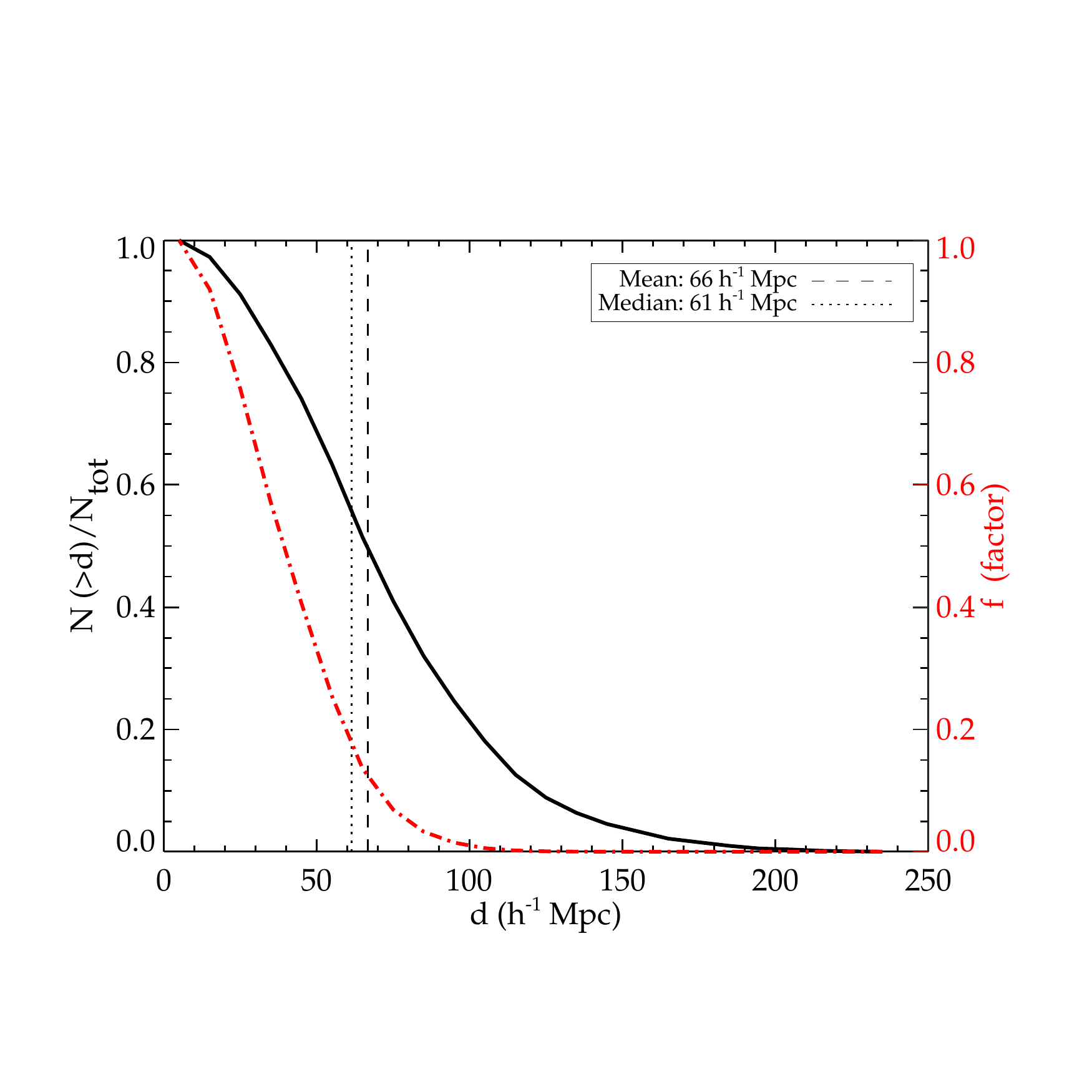}
\vspace{-1.9cm}

\caption{Normalized cumulative distribution of distances (black line) of the catalog of constraints. The dot-dashed red line is the same distribution to the power 3. It is used as a factor when improving the modeling of the peculiar velocity uncertainties. Dotted and dashed black lines are the median and mean distances of the catalog.}
\label{fig:factor}
\end{figure}

It is in a sense completely expected that the accuracy of the simulations drops with the distance from us: there are less data and they are more noisy and uncertain. However this systematic shrink of the most massive halos inherent to the technique should be accounted for to produce simulations reproducing the local clusters up to greater distances than those within the first 30\hMpc\ and above all to ensure a proper mass distribution within the local volume. A feature of the Wiener Filter and by extension of the Constrained Realization technique is to smooth the data when they are sparser and noisier. While this is absolutely correct when studying the sole reconstruction, the effect of smoothing when there are less data is undesirable for constrained simulations. Surely, the Constrained Realization technique compensates the effect by adding a random realization but unless this realization matches in a sense the observational data by imposing the same type of constraints at a given location, the chances of getting the clusters are drastically reduced with the distance from us. Indeed, the amount of data  in the catalog of constraints  drops drastically with the distance (see Figure \ref{fig:factor}).  

Since the Wiener Filter smooths also the data with respect to their uncertainties plus $\sigma_{NL}$ added in quadrature (see 2.1), it is possible to model differently this error term to account for the smoothing with the distance. The error term - uncertainties $\sigma$ (5\% after applying the bias minimization scheme) plus $\sigma_{NL}$ to sample a typical realization of the prior cosmological model - becomes after adding a third term:
\begin{equation}
\sigma'(d) = \sqrt{ \sigma(d)^2 + \sigma^{'2}_{NL}+  (f(d) \times \sigma_{eff})^2 } 
\label{eq:1}
\end{equation}
where $f$ is the red line factor given on Figure 1, it depends on the distance $d$ of the constraint whatever the direction of observation. This factor is the normalized cumulative distribution of distances in all directions to the power 3.

By default, $\sigma_{eff}$ is chosen to be equal to $\sigma_{NL}$ required to get $\chi^2$/dof$\sim$1 before adding the third term. $\sigma^{'}_{NL}$ can then be determined to get $\chi^2$/dof$\sim$1 after adding the third term. It appears that $\sigma^{'}_{NL}$ is not required anymore: $\sigma^{'}_{NL}$ =0 gives $\chi^2$/dof$\sim$1.  Note that alternatively, $\sigma_{eff}$ could be adjusted to get $\chi^2$/dof$\sim$1 so that in any case $\sigma^{'}_{NL}$ is not required anymore. Since there is no need to adjust $\sigma_{eff}$ here, this alternative is not further investigated. The modeling of the uncertainties as proposed above is thus similar to replacing $\sqrt{ \sigma^2 + \sigma_{NL}^2 }$ by $\sqrt{ \sigma^2 + (f \times \sigma_{eff})^2 }$ where $\sigma_{NL}$ used to be applied uniformly to the whole dataset. The purpose was to filter out the non-linear residual in the catalog of constraints but it did not take into account the filtering because of the decrease in the amount of data with the distance from us. Interestingly, required values of $\sigma_{eff}$ and $\sigma_{NL}$ necessary to get $\chi^2$/dof$\sim$1 are identical.

This $\sigma_{eff}$ is expected to force the Wiener Filter to smooth more the nearby datapoints than those further away to balance the opposite smoothing due to the decrease of the number of constraints with the distance. It does not affect the smoothing due to the noise and uncertainties directly linked to the distance (peculiar velocity) estimates.

One could argue that the number of data point is not isotropically fading with the distance. Still, \citet{2017arXiv170704267S} showed that the Wiener Filter is smoothing almost isotropically for the catalog used here. Adding this function of space in the factor $f$ does not improve the results presented in the next section for distant local clusters and degrades the very nearby clusters suggesting that this extra complexity does not constitute a solution to the weak anisotropy. Indeed, small fluctuations of $\sigma$ in the small inner volume because of a small anisotropy is not optimal for the Wiener Filter that works best on the large scales. In light of the above and since there is a priori no reason to use the anisotropic function on the large distance and not close by and since this anisotropy is almost inexistent, it is disregarded in the rest of the paper.

Subsequently, some tests were conducted on the mock catalogs developed in \citet{2015MNRAS.450.2644S} to further support this new modeling of the uncertainties. These mock catalogs, described in details in \citet{2015MNRAS.450.2644S}, reproduce the observational biases and the distribution of grouped data in the catalog of peculiar velocities. Before the modeling of uncertainties as proposed above, mass functions of re-simulations of the reference simulation obtained with the mock catalogs differ from that of the reference simulation at the large mass end: The reference simulation has on average 30\% more halos of masses above 10$^{14}$ h$^{-1}$~M$_\odot$ than the re-simulations. In other words the re-simulations lack 30\% of massive halos on average. This proves that the smaller mass functions at the high mass end found for the constrained simulations is not entirely due to the cosmic variance.  On the opposite, the uncertainty modeling described above permits recovering on average the mass function of the reference simulation at the large mass end in the new re-simulations: on average, the re-simulations have the same number of halos of masses greater than 10$^{14}$ h$^{-1}$~M$_\odot$ as the reference simulation. Such tests are in favor of modeling the uncertainties according to the fading of the constraints with the distance. Furthermore, any residual difference between the constrained simulations and the random simulations will then be a result of the cosmic variance.

The initial conditions for the 3$^{rd}$ generation of simulations constrained with the radial peculiar velocity catalog are then obtained in the standard way. Basically the steps are those described above (i to v) but in step (iii) the uncertainties $\sigma$ are replaced with those given by equation \ref{eq:1}. 

\section{Comparisons between the 3 generations}

Three sets of 9 simulations each sharing the same 9 random realization (in the constrained realization technique) are further used for the comparisons in that section. These sets are called 1$^{st}$, 2$^{nd}$ and 3$^{rd}$ generations. The difference between the 1$^{st}$ and 2$^{nd}$ generations has been described above in detail. It resides in the grouping scheme. The 3$^{rd}$ generation differ from the 2$^{nd}$ by the modeling of the uncertainties after the minimization scheme. The importance of this scheme has also already been discussed. Table \ref{Tbl:1} summarizes the different steps for the 3 generations as well as their relative importance. 

All these simulations are 500~\hMpc\ boxes run with 512$^3$ particles a resolution sufficient to study clusters at redshift zero (particle mass about 8$\times$10$^{10}\hmsun$, namely about 10$^3$ or more particles in objects larger than 10$^{14}~\hmsun$). They are run within the framework of Planck cosmology \citep[$\Omega_m$=0.307, $\Omega_\Lambda$=0.693, H$_0$=67.77\kms~Mpc$^{-1}$, $\sigma_8$~=~0.829,][]{2014A&A...571A..16P}. 

To perfect the comparisons, a set of 9 random simulations (sharing the same random realizations and settings) are also run.

\subsection{The local Large Scale Structure}

\begin{figure*}
\includegraphics[width=1 \textwidth]{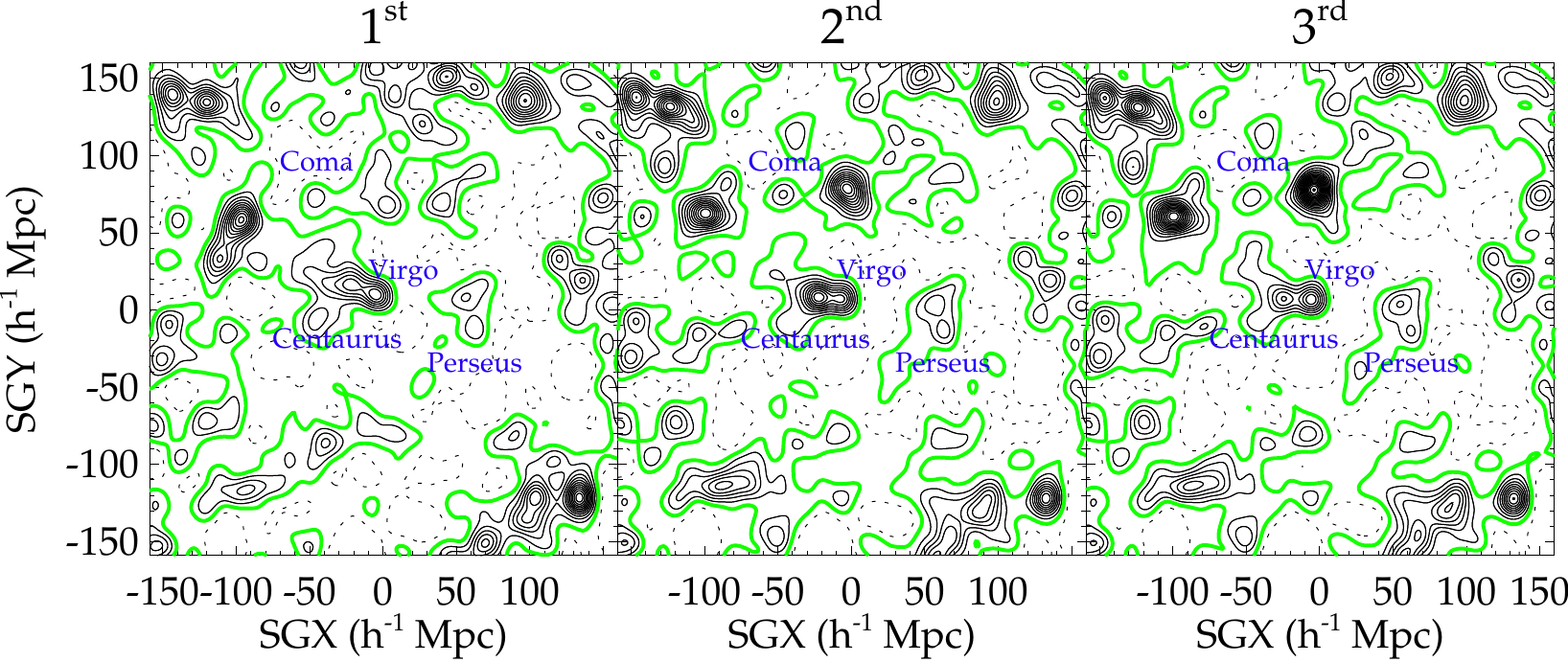} 
\caption{XY supergalactic slices of the local Large Scale Structure obtained in 3 constrained simulations sharing the same random realization. Different grouping schemes are used to remove non-linear motions from the constraint-catalog of galaxy radial peculiar velocities and a new modeling of the uncertainties permits accounting for the smoothing due to the decrease in the number of constraints with the distance. Left: Grouping scheme released with the catalog. Middle: Grouping scheme based on an advanced friend-of-friend algorithm. Right: Same grouping scheme but with the new uncertainty modeling presented in this paper. In all three cases, the bias minimization scheme is applied. The contours stand for the density field. Solid lines show overdensities while dashed lines represent underdensities. The green color is the mean field. A few structures are named in blue. Overall the local Large Scale Structure is properly reproduced in the three simulations but differences appear at the cluster scale. For instance, the overdensity at Coma's location is sharper and sharper from left to right.}
\label{fig:LSS}
\end{figure*}

Figure \ref{fig:LSS} shows the supergalactic XY slice of one simulation from each set of the three generations. The simulations share the same random realization. Solid (dashed) contours stand for the overdensities (underdensities). The green color delimits the overdensities from the underdensities. A few structures are identified with blue names. Overall the local Large Scale Structure is similar in the three cases confirming the efficiency of the constraining scheme on the large scales. However, differences appear at the cluster scale: from left to right, clearly the Coma region is getting denser and denser. The same goes for the Shapley concentration on the left side of Coma. Perseus is also a bit sharper from left to right. Only the densities for Virgo and Centaurus regions increase to decrease again. Since the masses of Virgo and Centaurus in the 2$^{nd}$ generation are slightly larger than observational estimates \citep{2018MNRAS.tmp..523S}, this is not in conflict with observations in a first qualitative approach. A detailed study of the masses in the next subsection confirms this assertion in a quantitative way.  

\begin{figure*}
\includegraphics[width=1 \textwidth]{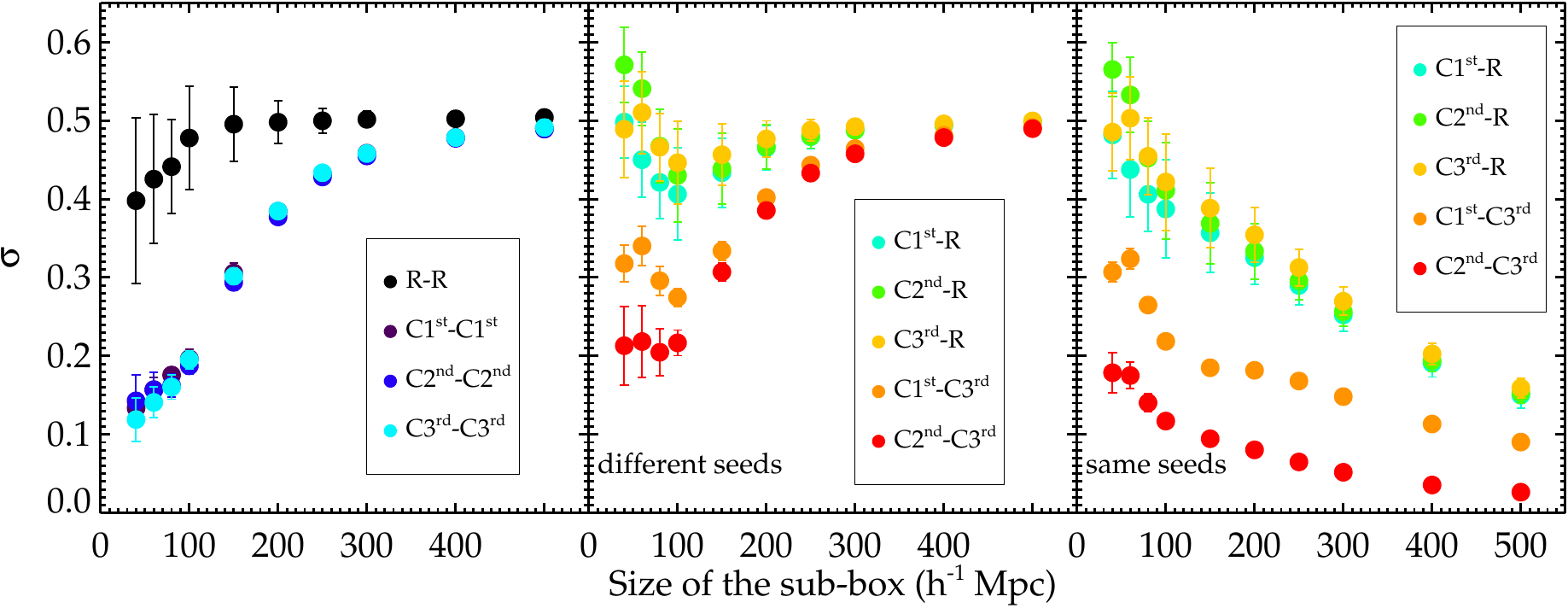}
\caption{Average variance (filled circle) and its standard deviation (error bar) between density fields of simulations as a function of the size of the compared sub-box. From left to right: comparisons of pairs of random (R, black) and constrained (C${\rm 1^{st}}$ violet, C${\rm 2^{nd}}$ dark blue, C${\rm 3^{rd}}$ light blue) simulations, comparisons between random and constrained simulations (light blue, green and yellow) as well as between constrained simulations obtained with different groupings or uncertainty modelings (orange and red) that do not share the same random realization (middle) and that share the same random realization (right). The goal of the constrained simulations is fulfilled: the cosmic variance is reduced with respect to that of random simulations.}
\label{fig:cosvar}
\end{figure*}

To enhance these qualitative observations, cell-to-cell comparisons between pairs of simulations are conducted. First, cells are compared within the full box. The scatter around the 1:1 relation is derived. Once all the scatters are obtained for a given type of simulation pairs, their mean and variance are computed. Second, because simulations are known to be more constrained in the center of the box where most of the constraints are, cells are compared only in sub-boxes. All the resulting mean scatters (as defined above when comparing the full boxes and different size sub-boxes) and their variances are reported in Figure \ref{fig:cosvar} as a function of the size of the sub-boxes within which cells are compared between simulations. 

The left panel of the figure gives the cosmic variance of the three generations of constrained simulations (C1$^{st}$, C2$^{nd}$, C3$^{rd}$) as well as that of the random simulations (R). As already shown for the two first generations \citep{2016MNRAS.455.2078S,2018MNRAS.tmp..523S}, the cosmic variance is decreased by a factor 2 to 3 within the inner part of the box: the scatters obtained when comparing pairs of constrained simulations of the same nature are considerably smaller than those obtained when comparing random simulations. Interestingly, for the 3$^{rd}$ generation of constrained simulations presented in this paper, the cosmic variance in the inner part of the box is even smaller than for the two other generations suggesting a slightly better control of the constrained region for that last generation. 

The two other panels of the figure show the cosmic variance between simulations of different types be they constrained and random or constrained but of different generations. The comparisons between the 1$^{st}$ and 2$^{nd}$ generations has already been led by \citet{2018MNRAS.tmp..523S}. They are thus not reiterated here. In the middle panel, mean scatters and their variances are derived only with pairs of simulations that do not share the same random realization. On the contrary, the right panel gives the mean scatters and their variances only for pairs of simulations sharing the same random realization.

\begin{figure}
\vspace{-2cm}

\includegraphics[width=0.5 \textwidth]{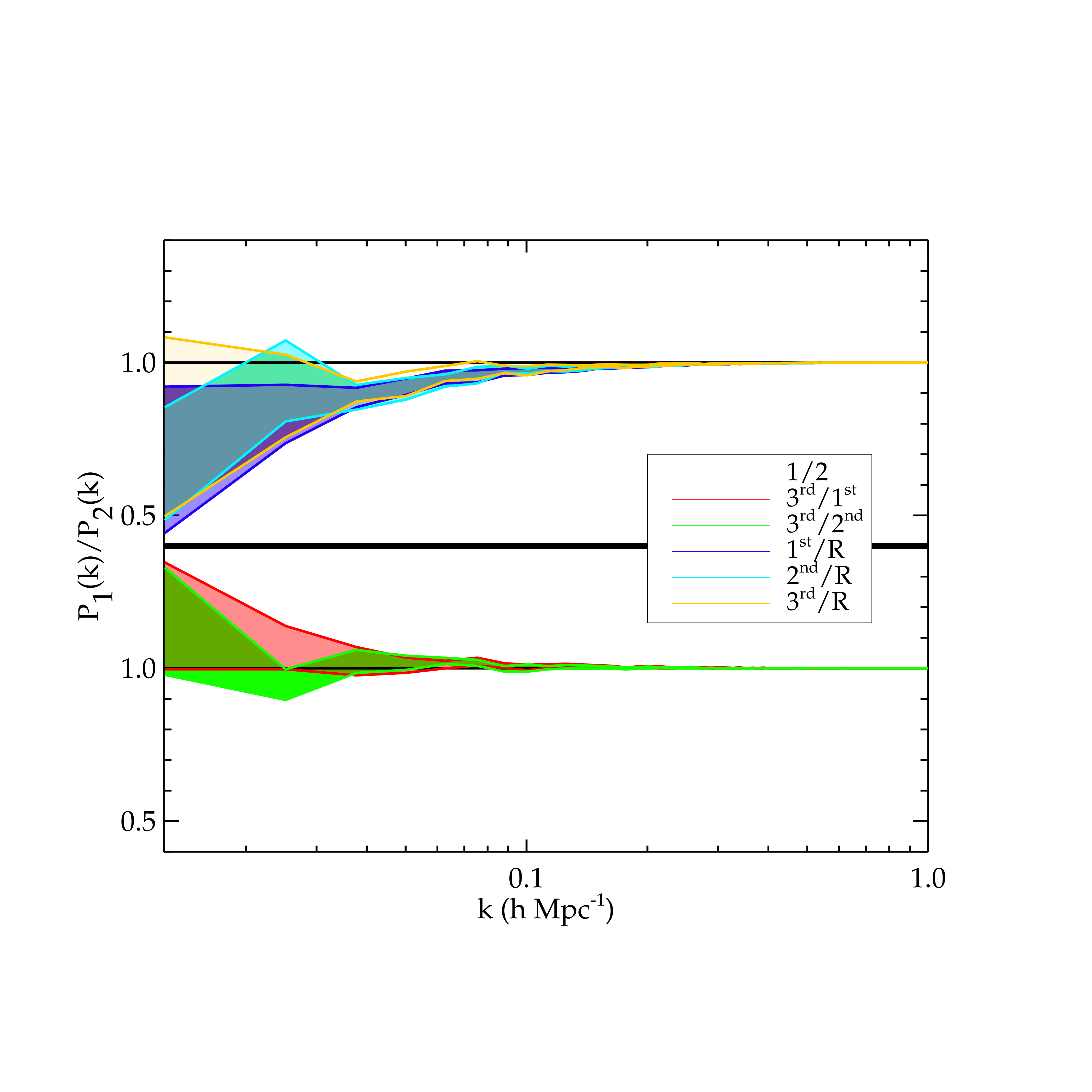}
\vspace{-3.2cm}

\includegraphics[width=0.5 \textwidth]{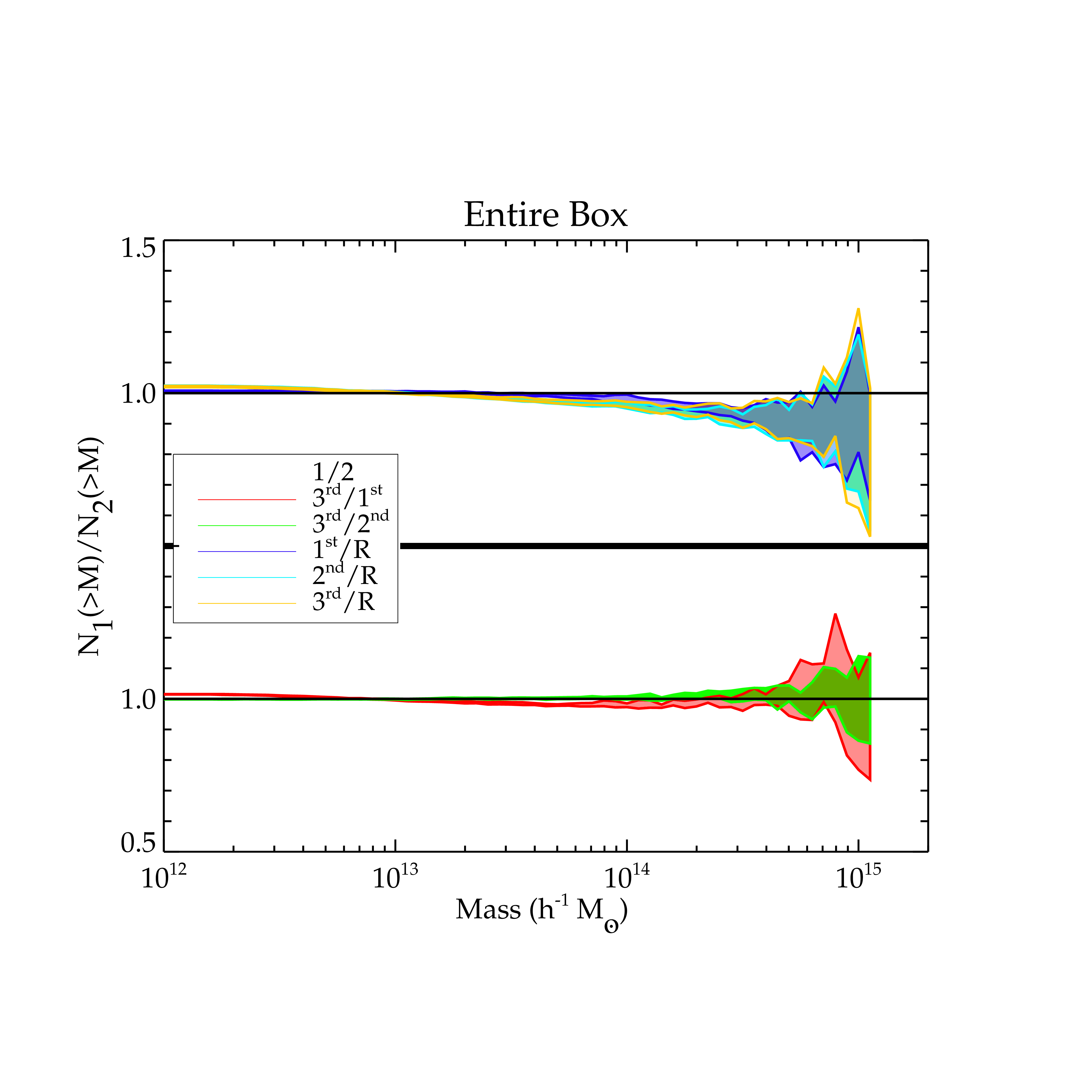}

\vspace{-2.8cm}
\includegraphics[width=0.5 \textwidth]{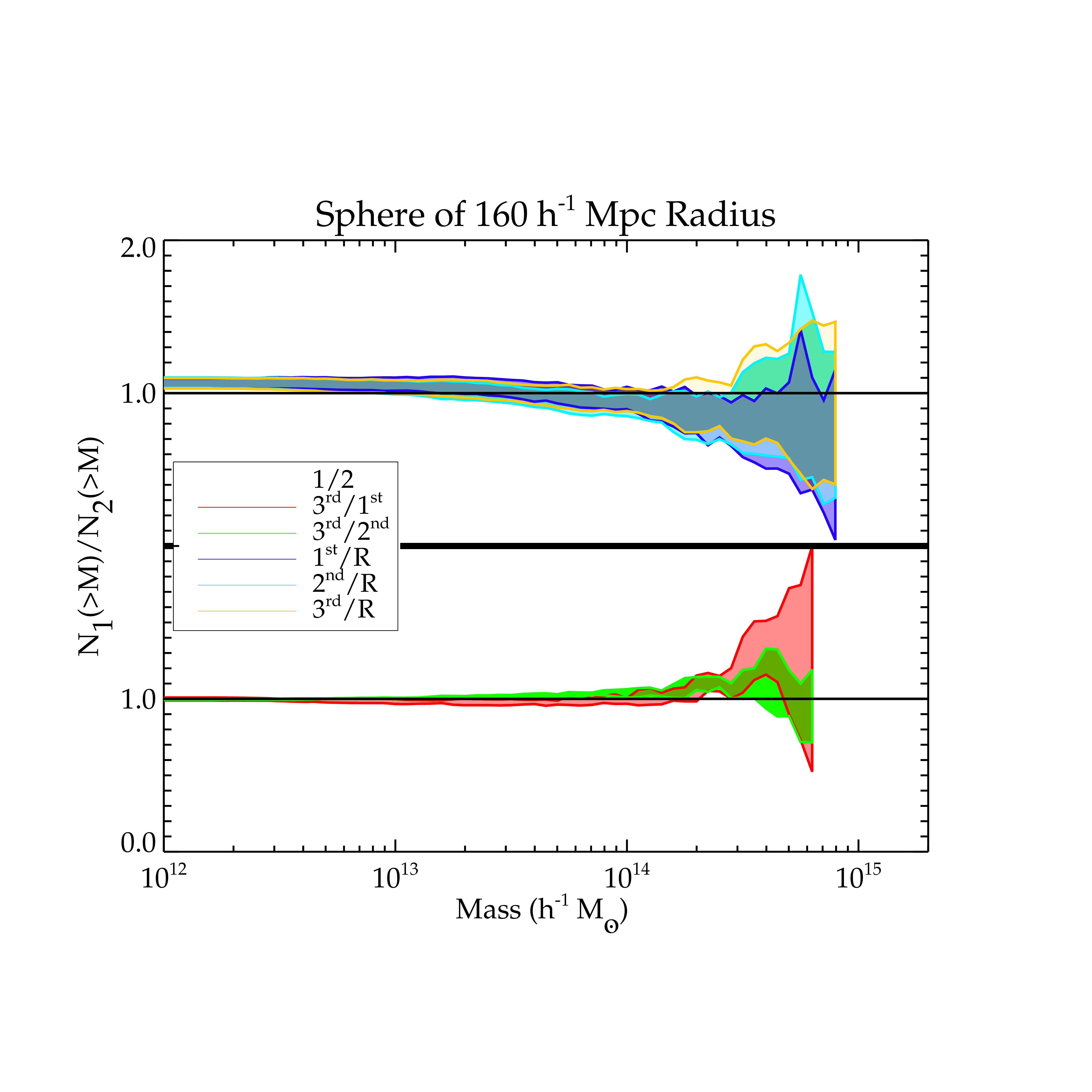}
\vspace{-1.5cm}

\caption{Top: 1$\sigma$ confidence interval of the ratio of the power spectra of constrained and random simulations (blue and yellow) and of constrained simulations (red and green). Middle and bottom: the same as the top panel but for the mass functions of the entire box and of a 160 \hMpc\ radius sphere.}
\label{fig:powspecmass}
\end{figure}

Comparisons betwen pairs of simulations that do not share the same random realization (middle panel) show that overall constrained simulations differ in a similar way from the random simulations (light blue, green and yellow). There are small differences in the inner part of the box where the 2$^{nd}$ and 3$^{rd}$ generations simulations differ more from the random simulations than those of the 1$^{st}$ generation. Interestingly, the 2$^{nd}$ generation simulations versus the random simulations present the highest scatter in the inner 50~\hMpc\  while the 3$^{rd}$ generation simulations versus the random ones do so at larger distances. Within the inner box (30~\hMpc) the latter have the same scatter as the 1$^{st}$ versus random simulations. It is a first hint at a better repartition of the mass in the local Large Scale Structure between the different clusters for the 3$^{rd}$ generation simulations. Indeed, the 2$^{nd}$ generation simulations have very nearby clusters slightly overmassive while further clusters are undermassive. This can explain the larger scatters observed in the very central part of the box: larger than expected overdensities to be compared to overall voids in random simulations - higher probability of finding a void in a small sub-box in a random realization than in a constrained simulation since structures are there by construction. The 1$^{st}$ generation simulations have undermassive clusters everywhere except the Virgo cluster that is in good agreement with observations. Hence the smaller scatters. As for the 3$^{rd}$ generation simulations versus the random simulations, since the mass is better distributed in the constrained simulations, the scatters are overall of the same order in all the sub-boxes. Comparisons between the different generation simulations confirm the impact of both the grouping scheme and the new modeling of the uncertainty in the inner part of the box: beyond 160~\hMpc, the measured scatters are similar as those obtained when comparing simulations of the same generation (left panel). However, within the 160~\hMpc, the scatters are larger when comparing simulations of different generations (middle panel) than those of the same generation (left panel).

Finally, the right panel gives the variance between pairs of simulations that share the same random realization. As expected, when considering the full box, since the non constrained regions overcome the constrained ones, the scatters are the smallest. They then increase when considering smaller and smaller sub-boxes. In other words, the random realization dominates in the whole volume thus comparisons involve almost the same fields. However, within the inner 160~\hMpc\ part of the box, the scatters have the same values as those obtained when comparing simulations that do not share the same random realization. Overall the scatters follow the same order: they increase from the comparisons between the simulations of the 2$^{nd}$ and 3$^{rd}$ generations to the comparisons between the 3$^{rd}$ generation and random simulations. The same conclusions as before can thus be drawn. \\

Another quantitative analysis can be led by comparing the power spectra and the mass functions of all these simulations. To derive the mass functions, the dark matter halos are identified with Amiga's halo finder \citep{2009ApJS..182..608K}. Figure \ref{fig:powspecmass} presents the ratios of all these power spectra (top panel) and mass functions for the entire boxes (middle panel) and for the 160~\hMpc\ radius spheres (bottom panel). The first panel of the figure shows that the power spectra of the constrained simulations are overall below those of the random simulations (mean of the yellow and blue areas below 1) as already observed by \citet{2016MNRAS.455.2078S}. However, the 2$^{nd}$ and 3$^{rd}$ generation simulations have slightly larger power spectra. Mostly on the large scales ($>$100~\hMpc) for the 2$^{nd}$ generation simulations while it is better distributed on all scales above about 30~\hMpc\ for the 3$^{rd}$ generation (mean of the red area above 1 for k$<$0.03~\hMpc). This is overall a good result since the probability of having such a power spectrum given a prior (Planck) for the local Universe is increased from the 1$^{st}$ to the 3$^{rd}$ generation.   

Regarding the mass functions, the same observations can be made. Clearly the 2$^{nd}$ and 3$^{rd}$ generation simulations have more massive halos than the 1$^{st}$ generation ones. This affirmation is more visible within the 160~\hMpc\ radius sphere. The red area has clearly a mean above 1 at the high mass end. Namely, the 3$^{rd}$ generation simulations have more massive haloes than the 1$^{st}$ generation simulations. With respect to the 2$^{nd}$ generation simulations, the more massive halos are better distributed over the whole high mass end in the 3$^{rd}$ generation simulations: the mean green area is slightly above 1 for masses greater than 10$^{14}~\hmsun$ but below 1 at the extreme end ($\sim$10$^{15}~\hmsun$). 
These observations are again in favor of a better distribution of the mass within the local volume.

To conclude this subsection, the local Large Scale Structure is overall reproduced by the simulations in all three generations. However, the 2$^{nd}$ then 3$^{rd}$ generation simulations have more probable power spectra with respect to Planck's. Additionally they have more massive halos. The 3$^{rd}$ generation simulations are the most probable given the prior cosmological model and have a better repartition of the mass within the local clusters. The next subsection will shade some lights on this assertion by studying in detail the halos found in the simulations.

\subsection{Local galaxy clusters}

\begin{table}
\begin{center} 
\begin{tabular}{lrrrrl}
\hline \hline
(1) & (2) & (3) & (4) & (5)  \\
Cluster & sgl & sgb & d & M\\
 & ($^\circ$) & ($^\circ$) & (Mpc) & (10$^{14}$~M$_\odot$) \\
\hline
Virgo 	& 103.0008 & -2.3248  &14.9 & 7.01\\
Centaurus & 156.2336 & -11.5868  & 38.7 & 10.8 \\
 Hydra  & 139.4478 &	 -37.6063 & 41.0  & 4.39 \\ 
Perseus  &347.7159 	& -14.0594 &  52.8 &	16.3 \\
Coma & 89.6226 & 8.1461 & 73.3  &15.9  	\\
\hline
\hline
\end{tabular}
\end{center}
\vspace{-0.25cm}
\caption{Clusters from \citet{2015AJ....149..171T} with H$_0$=75~km~s$^{-1}$~Mpc: (1) cluster name, (2) supergalactic longitude, (3) supergalactic latitude, (4) distance, (5) virial mass.}
\label{Tbl:2}
\end{table}

\begin{figure*}
\includegraphics[width=0.95 \textwidth]{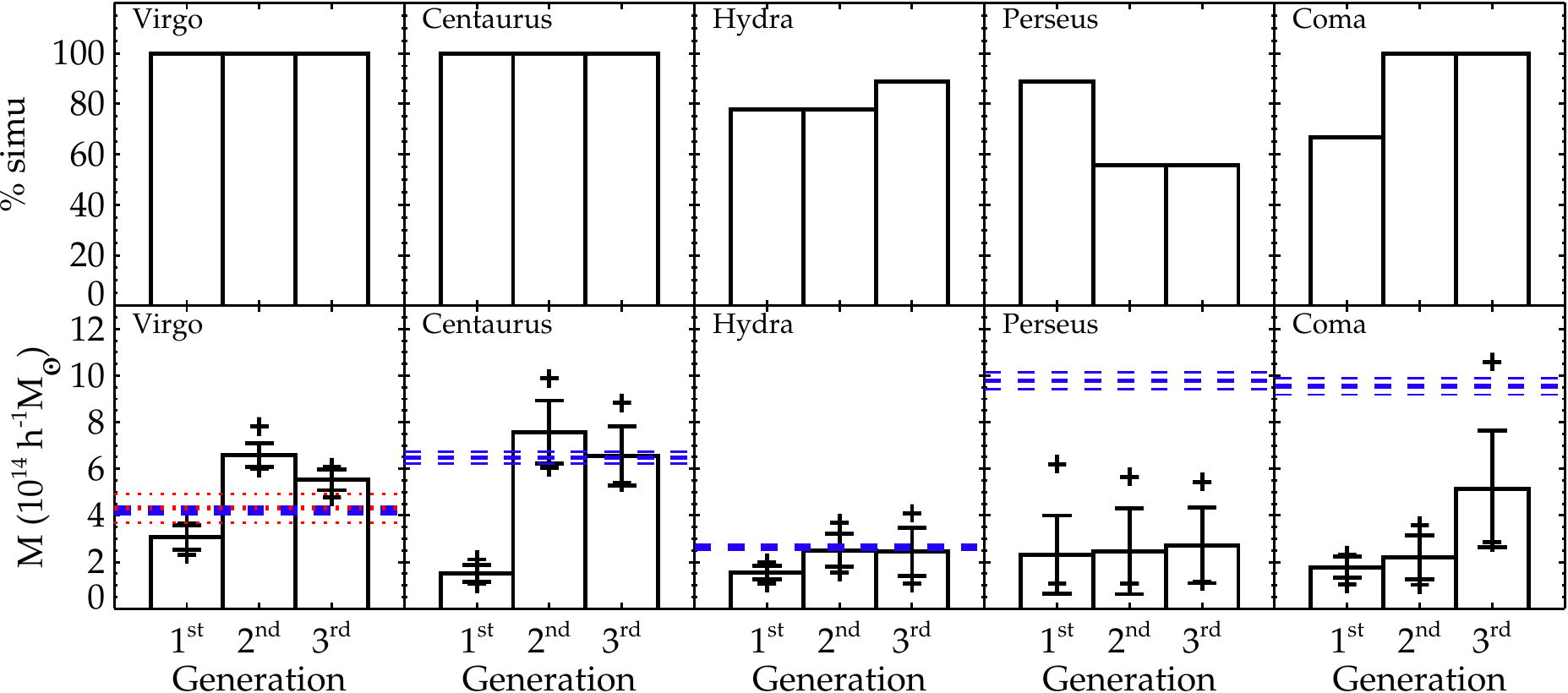} 
\caption{Top row: percentage of simulations with a simulacrum of the observed cluster whose name is given in the top right corner of each panel for the 3 generations of constrained simulations. Bottom row: average mass (histogram), standard deviation (error bar) and minimum / maximum masses (crosses) of the different simulacra for the 3 generations. Thick blue dashed and red dotted lines are observational mass estimates. The thiner dashed and dotted lines stand for the uncertainty on the proportionality factor between observational masses and M$_{200}$. Without any doubt, the 3$^{rd}$ generation simulations are those that host overall the best simulacra of these local clusters.}
\label{fig:clusters}
\end{figure*}

Simulacra of 5 of the local clusters (Virgo, Centaurus, Hydra, Coma and Perseus) are looked for in the three sets of constrained simulations. A dark matter halo is considered as a simulacrum of a local observed cluster if it is at the proper location with respect to observational estimates (observed and simulated distances cannot differ by more than 30\%) and if its mass is above 10$^{14}~\hmsun$. It is important to note that if simulacra are slightly shifted in positions with respect to the observed cluster, their shifts are consistent between each other. Namely simulacra are only a few megaparsecs away from each other from one simulation to another. The top row of Figure \ref{fig:clusters} gives the probability of finding a simulacrum for a given cluster in each set of constrained simulations. A unique Virgo (Centaurus) simulacrum is found in each one of the simulation, whatever the generation is. A Coma look-alike is however found in only 65\% of the 1$^{st}$ generation simulations while it is in a 100\% of the other generation simulations. The probability of finding a Hydra simulacra is increased up to 90\% in the 3$^{rd}$ generation simulations compared to the 80\% before. Perseus is the only local cluster whose probability decreases in the 2$^{nd}$ and 3$^{rd}$ generations. It is worth noting that this is only due to the cut in mass at 10$^{14}~\hmsun$. Would the cut be slightly smaller then the probability in the 2$^{nd}$ and 3$^{rd}$ generation simulations would increase since the halo candidates that are rejected are barely below 10$^{14}~\hmsun$ in these cases. In addition, the currently used catalog of constraints does not have much datapoints in that region. With the third catalog of constraints that contains more distances in that region \citep{2016AJ....152...50T} better results can be expected for that particular area. Globally, the 3$^{rd}$ generation of simulations increases the probability of finding simulacra of the local clusters.

The bottom row of that same figure is even more important as it shows the mean, standard deviation, minimum and maximum masses, M$_{200}$ (i.e. the mass enclosed in a sphere with a mean density of 200 times the critical density of the Universe), of these dark matter simulacra. Comparing these M$_{200}$ to observational estimates is not direct. For instance \citet{2015AJ....149..171T} published recently the virial masses of these local clusters in M$_\odot$ to be compared with M$_{200}$. These latter are proportional to the virial mass (given by the halo finder) via a factor of 0.80$\pm$0.03 \citep[e.g.][]{2016MNRAS.460.2015S}. Assuming the virial masses given by both observational estimates (cf.  Table \ref{Tbl:2}) and the halo finder to be roughly similar, the former are shown for comparisons on Figure \ref{fig:clusters} with thick blue dashed lines as well as the 1$\sigma$ uncertainty of the conversion factor with thinner lines. The red dotted lines stand for a very recent estimate of Virgo's mass (M) via the first turn around radius by \citet{2017ApJ...850..207S} with M$_{200}$=0.70$\pm$0.10~M \citep{2016MNRAS.460.2015S}. Again, only the uncertainties on the conversion factor are retained for the plot. 

Virgo and Centaurus dark matter halos are more massive than in the 1$^{st}$ generation simulations but less massive than in the 2$^{nd}$. This is in excellent agreement with observational estimates shown as red and/or blue lines. While masses of Virgo simulacra are compatible with the observational estimates for all three generations within 3$\sigma$s, they are within 2$\sigma$s only for the 1$^{st}$ and 3$^{rd}$ generations. As for Centaurus like halos clearly undermassive in the 1$^{st}$ generation, their average mass is now clearly on top of the observational estimates. The slightly smaller masses in the 3$^{rd}$ generation with respect to those in the 2$^{nd}$ generation are even in better agreement with observations. Coma simulacra have more than doubled (almost tripled) their masses. The 3$^{rd}$ generation simulations are the only ones hosting dark matter halos that are within 2$\sigma$s of the observational estimate. Hydra simulacra are in perfect agreement with observations for both the 2$^{nd}$ and 3$^{rd}$ generation simulations. Although this agreement is smaller the 1$^{st}$ generation simulations, masses are still within 3$\sigma$s. Perseus-like halos are the only ones that are not drastically improved. Although the mean mass increases from the 1$^{st}$ to the 3$^{rd}$ generation simulations, they are still barely within 3$\sigma$s and more within 4$\sigma$s. As explained before, this region is not well probed by the catalog of constraints used here.

Regardless, without any doubt, the 3$^{rd}$ generation simulations are those that host overall the best simulacra of the local clusters. The drastic increase in mass of the very nearby cluster-like at the expense of the farther local clusters visible in the 2$^{nd}$ generation simulations is erased by the new modeling of the uncertainties.\\

\begin{figure*}
\includegraphics[width=0.95 \textwidth]{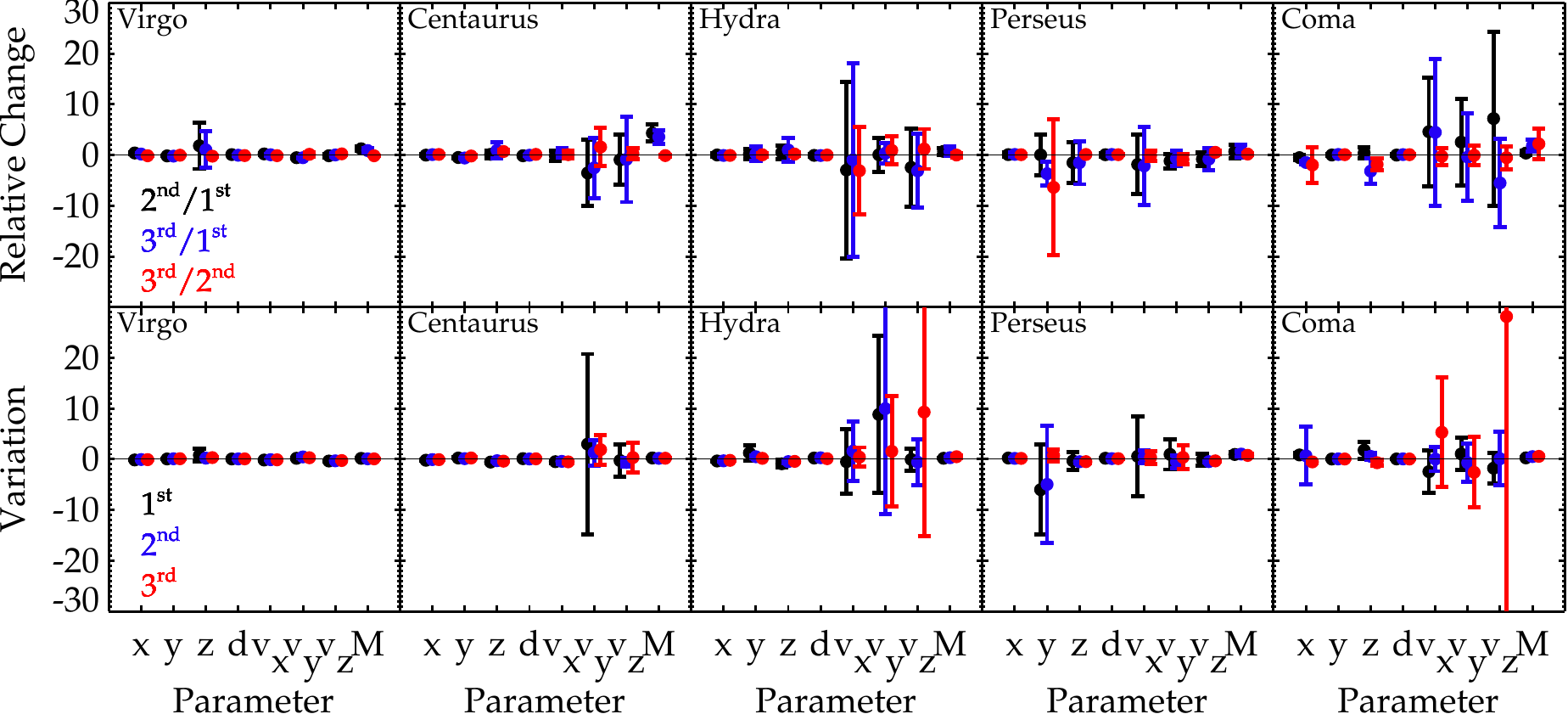} 
\caption{Top row: relative change and standard deviation (filled circle and error bar)  between the parameters of the cluster simulacra obtained in the simulations produced with the different grouping schemes or uncertainty modeling. Bottom row:  variation (filled circle) and standard deviation (error bar) of the parameters of the cluster simulacra found in the 1$^{st}$ (black), 2$^{nd}$  (blue) and 3$^{rd}$ (red) generations of constrained simulations.}
\label{fig:clusprop}
\end{figure*}

The top (bottom) row of Figure \ref{fig:clusprop} probes the relative change (variation) between simulacra hosted by different (the same) generation simulations. Virgo look-alike is incredibly stable between the different generations (top row) and within a given generation (bottom row). The only notable difference is in the $z$ direction. This direction is more constrained in the 2$^{nd}$ and 3$^{rd}$ generation simulations than in the 1$^{st}$ generation ones. This is remarkable since this is the less constrained direction because of the zone of avoidance due to the Milky Way's dust. Centaurus simulacrum becomes also quite stable. The only exceptions being the $y$ and $z$ components of the velocities. Still the $y$ component is much more stable in the 2$^{nd}$ and 3$^{rd}$ generations than it was before. Dark matter halos that are substitutes of Coma and Hydra are more stables in terms of positions in the 2$^{nd}$ and 3$^{rd}$ generation simulations. As for Perseus-like, its velocity components are more stable for the 2$^{nd}$ and 3$^{rd}$ generation simulations, especially the $x$ component. Remarkably, its position is more stable only in the 3$^{rd}$ generation simulations, especially the $y$ component.

To conclude the biggest impact of the new modeling of the uncertainties is to reach better agreement between observed and simulated masses.

\section{Conclusion}

This paper investigates the method used so far to get local Universe like simulations that resemble the local Universe down to the cluster scales using only peculiar velocities (no densities) as constraints. Namely not only do these simulations look like the local Large Scale Structure but they also host dark matter halos that can be taken as simulacra of local clusters. So far the simulations obtained with the method described in this paper from the catalog of distances to the constrained realization technique going through a series of steps all more important than each others - grouping, minimization of biases, reverse Zel'dovich approximation, replacement of radial by 3D velocities - permitted getting great simulacra for the Virgo and Centaurus galaxy clusters. Local clusters further away seemed irremediably too faint without adding any additional density constraints. 

However, point-like density constraints are not optimal: 1) mass estimates of clusters are not easily obtained and subject to uncertainties~; 2) chosen mass values can bias the statistical results obtained with the resulting constrained simulations. Simulations constrained with both types of constraints cannot cover as much ground as simulations constrained solely with peculiar velocities. For instance, the mass of a cluster induced by a given environment is not deductible anymore since it has been independently constrained ; 3) its is impossible to add a density constraint for all the clusters especially for those in the zone of avoidance ; 4) adding a full redshift survey to the peculiar velocities is complicated by the fact that both surveys must be greatly consistent to predict the same field (positions included) down to the cluster scale. In light of the above, it seems grandly recommended to simulate well known observed clusters using solely radial peculiar velocities not to bias the results and conclusions for these clusters but also for other clusters in the local volume.

Observing that the fading of the cluster masses in the constrained region of the simulations is correlated to the decrease in the amount of data in the catalog with the distance from us (the center of the box), a new modeling of the peculiar velocity uncertainties after the minimization scheme is proposed. This new modeling is built to account for the smoothing of the reconstruction technique due to the decreasing amount of data with the distance. It does not affect the filtering of the data due to their uncertainties inherent to their measurements. This new modeling appears similar to replacing the uncertainty by a sum in quadrature of this same uncertainty and a factor $f$ times a $\sigma$ effective. The factor $f$ depends on the distance of the constraint from the observer. It is obtained with the normalized distribution function of the constraints with the distance to the power 3. 
The $\sigma$ effective is set by default to the non-linear $\sigma_{NL}$ value required, before adding $\sigma$ effective, such that the data are sampling a typical realization of the assumed prior power spectrum model in the linear theory ($\chi^2$/dof$\sim$1). After adding $\sigma$ effective, $\sigma_{NL}$  appears not to be required anymore ($\sigma_{NL}$=0).

Multiple comparisons between the previous generation simulations to those obtained with this new modeling of the uncertainties reveal that the latter have overall a power spectrum with a higher probability given the prior cosmological model (Planck) and that they have a better distribution of the mass within the local Large Scale Structure between the different clusters. Consequently, although both the previous and new generation simulations overall reproduce the local Large Scale Structure, the latter generation permits producing simulacra of local clusters further away than Virgo and Centaurus such as the Coma cluster with appropriate masses.

Zoom-in simulations of the Coma cluster and of a larger volume than the 30\hMpc\ radius inner sphere become now possible to study local clusters and their effects. Mapping the impact of the local Universe on distant observations can follow.

\section*{Acknowledgements}
JS acknowledges stimulating discussions with her collaborators: Klaus Dolag, Stefan Gottl\"ober and Yehuda Hoffman. The author would like to thank the referee for comments that helped improve the manuscript. JS acknowledges support from the Centre National d'\'etudes spatiales (CNES) postdoctoral fellowship program as well as from the ``l'Or\'eal-UNESCO Pour les femmes et la Science'' fellowship program. The author gratefully acknowledges the Gauss Centre for Supercomputing e.V. (www.gauss-centre.eu) for providing computing time on the GCS Supercomputer SuperMUC at LRZ Munich.


\bibliographystyle{mnras}

\bibliography{biblicomplete}
 \label{lastpage}
\end{document}